\documentclass[twocolumn,showpacs,preprintnumbers,amsmath,amssymb]{revtex4}

\usepackage{psfig}
\usepackage{graphicx}
\usepackage{dcolumn}
\usepackage{bm}
\usepackage[dvips]{color}

\newcommand{\be}{\begin{equation}}
\newcommand{\ee}{\end{equation}}

\newcommand{\op}[1]{\hat{#1}}
\newcommand{\mean}[1]{\langle #1 \rangle}

\begin{document}
\title{Radiative emission dynamics of quantum dots in a
single cavity micropillar}

\author{M. Schwab}
\author{H. Kurtze}
\author{T. Auer}
\author{T. Berstermann}
\author{M. Bayer}
\affiliation{Experimentelle Physik II,
             Universit\"at Dortmund,
             44221 Dortmund, Germany}

\author{J. Wiersig}
\author{N. Baer}
\author{C. Gies}
\author{F. Jahnke}
\affiliation{Institute for Theoretical Physics,
             University of Bremen,
             28334 Bremen, Germany}

\author{J.~P. Reithmaier}
\author{A. Forchel}
\affiliation{Technische Physik,
             University of W\"urzburg,
             Am Hubland,
             97074 W\"urzburg, Germany}

\author{M. Benyoucef}
\author{P. Michler}
\affiliation{5th Physics Institute
             University of Stuttgart,
             70550 Stuttgart, Germany}

\date{\today}

\begin{abstract}
The light emission of self-assembled (In,Ga)As/GaAs quantum dots
embedded in single GaAs-based micropillars has been studied by
time-resolved photoluminescence spectroscopy. The altered
spontaneous emission is found to be accompanied by a
non-exponential decay of the photoluminescence where the decay
rate strongly depends on the excitation intensity. A microscopic
theory of the quantum dot photon emission is used to explain both,
the non-exponential decay {\sl and} its intensity dependence. Also
the transition from spontaneous to stimulated emission is studied.
\end{abstract}


\maketitle

\section{\label{intro}Introduction}

The possibility of altered spontaneous emission by modifying the
photonic environment, known as the Purcell effect
\cite{Purcell46}, allows one to tailor the optical emission
properties of quantum dots (QDs). Spontaneous emission is caused
by fluctuations of the vacuum electromagnetic field, so that its
change represents a true quantum effect. The photonic environment
can be altered by modifying the density of optical modes, to which
the QD electronic transitions can couple, and/or by modifying the
amplitude of the vacuum field at the QD location. Both changes
have been achieved by placing the QDs in a resonator structure
with size of the order of the light wavelength, in which the
electromagnetic field is three-dimensionally confined
\cite{Gayral99,Gerard98,Graham99,Bayer01,Solomon01,Smith99}. As a
consequence the mode spectrum becomes discretized, and the vacuum
field amplitude can be significantly modified.

The light emission of QDs in optical cavities has been a very
active field of solid state research during recent years. The
altered spontaneous emission dynamics of QDs has been demonstrated
using different resonator types, such as microdisk structures
\cite{Gayral99}, patterned cavity pillars
\cite{Gerard98,Graham99,Bayer01,Solomon01} or photonic crystal
defects \cite{Smith99}.  Its experimental verification requires
time-resolved photoluminescence (PL) measurements. Since QDs are
often considered as artificial atoms, it became a standard in
these studies to use the exponential decay known from two-level
emitters and to carry it over to a QD system in order to quantify
the emission dynamics and in particular the Purcell effect. Then
the decay time $\tau$ of the emitter in the presence of the cavity
follows from
\begin{eqnarray}
  \frac{\tau_{0}}{\tau}=\frac{2}{3} F_P
  \frac{|\vec{E}(\vec{r})|^{2}}{|\vec{E}_{max}|^{2}}
  \frac{\Delta \lambda^{2}_c}{4(\lambda_e - \lambda_c)^{2} + \Delta
  \lambda^{2}_{c}} \cos ^2 \vartheta +\frac{\tau_{0}}{\tau_{leak}},
\label{eq:lifetime}
\end{eqnarray}
where $\tau_0$ is the decay time in a spatially homogeneous
medium, which is determined by the Weisskopf-Wigner decay rate
\cite{Meystre}. The second term on the right hand side models the
emission into leaky modes. The first term describes the QD
emission at wavelength $\lambda_e$ into a cavity mode at
wavelength $\lambda_c$. An emitter at location $\vec{r}$ is
subject to an electric field $\vec{E}(\vec{r})$ whose amplitude
varies between the maximum value $|\vec{E}_{max}|$ in a field
antinode and zero for a node position. $\vartheta$ is the angle
between the electric field vector and the dipole moment of the
electronic transition.

The Purcell factor $F_P$ gives the enhancement of the emission
decay rate in the resonator in comparison to the homogeneous
medium,
\begin{eqnarray}
  F_P = \frac{3 \lambda_e^3}{4 \pi^2 n^3} \;
  \frac{ g\,Q}{V_c}.
\end{eqnarray}
Here, $Q$ is the quality factor of the resonator, $V_c$ is the
effective mode volume in the cavity  with refractive index $n$,
and $g$ is the mode degeneracy. The application of Eq.~(1)
requires that the emitter linewidth $\Delta \lambda_e$ is much
smaller than the cavity linewidth $\Delta \lambda_c$. This is well
fulfilled for QDs at cryogenic temperatures.

A closer inspection of the literature reveals, however, that in
many cases a non-exponential decay of the time-resolved PL is
observed for a wide variety of QD resonator systems
\cite{Lodahl04,Wang01,Cao02}, and even for QDs without optical
cavities \cite{Krest02,Oliver03}. While this  in itself
complicates the quantification of the altered spontaneous emission
in terms of a constant decay time, we additionally report a strong
dependence of the time-resolved PL decay on the excitation
intensity, also far below the stimulated emission regime. The
decay rate continuously increases from the weakest possible
pumping, for which we can detect a PL signal, up to the laser
threshold. Our experimental results are obtained by time-resolved
PL measurements using (In,Ga)As/GaAs QDs in GaAs-based pillar
microcavities. In these resonators, two distributed Bragg
reflector (DBR) mirrors provide the optical confinement along the
vertical direction. The patterning in the form of pillars provides
an additional efficient mode confinement in transverse direction
due to total internal reflection.  Using micro-PL, the emission of
individual pillars can be analyzed after excitation of the QDs
with a short laser pulse.

In the past, various effects have been proposed to explain the
frequently observed non-exponential decay of the PL under these
conditions: For example, the experimental data were recorded from
a QD ensemble.  As the emitter locations vary inside the
resonator, for each position a different decay rate is expected
from Eq.~(1). Hence, the integrated intensity measured in the
experiment does in general not exhibit a mono-exponential decay.
In addition, the ensemble exhibits fluctuations in the QD emission
energies and dipole matrix elements, which might lead to
deviations from a single exponential decay. In our calculations,
these effects were included. For the studied situation, they
influence only weakly the shape of the time-resolved PL signal.
Additional experimental evidence that the different positions in
the cavity and variations of the QD emission energies are not the
prime reason for the non-exponential character of the decay can be
obtained from studies of the time-resolved PL signal for QDs
without microcavity, which is otherwise not the subject of this
paper. \cite{Auer06} While the microcavity enhances the
non-exponential shape of the decay, also without microcavity the
effect can be observed.

As a further alternative, coupling of bright and dark exciton
states via spin-flip processes has been suggested as an origin for
the non-exponentiality \cite{Bayer02,PattonPRB03}. We discuss in
detail below that this mechanism can be ruled out for the present
experiments as well. Furthermore, none of the alternative
mechanisms can explain the observed strong dependence on the
excitation intensity.

Based on a microscopic theory for QD carriers interacting with the
quantized light field, we analyze the time-resolved PL. The
underlying set of equations is referred to as semiconductor
luminescence equations (SLE), which have been used in the past for
quantum well systems \cite{Jahnke1998}. While the PL of an (ideal)
ensemble of two-level emitters shows an exponential decay with a
time-constant independent of the excitation conditions, this is
found to be very different for semiconductor QDs. The
recombination of an excited electron requires the presence of a
hole. Scattering and dephasing processes reduce the correlation
between optically generated electron-hole pairs and, thus, lead to
a distinct departure from the simple two-level picture
(corresponding to independent excitons with fully correlated
electron-hole states). Our calculations of the electron-hole
recombination reveal the intrinsic nature of the non-exponential
decay and its excitation intensity-dependence. While other
(extrinsic) processes could be important in other specific
experiments, they do not show the discussed intensity dependence
and are limited to special systems or excitation conditions.

The article is organized as follows. In the next section we
discuss the samples under study as well as the experimental
technique. In Section 3 we present the experimental results and
give a preliminary analysis. In Section 4 the theoretical model is
introduced and the relation to the experiment is discussed in
detail. The article is concluded by a summary.

\section{\label{sample} Sample and Experiment}

The planar microcavity sample was grown by molecular beam epitaxy
 on a (100)-oriented undoped GaAs substrate, with a GaAs
buffer layer of 0.4~$\mu$m thickness. The GaAs $\lambda$-cavity
layer was sandwiched between two DBRs, consisting of 23 and 20
alternating AlAs/GaAs films for the bottom and the top mirrors,
respectively. Each film is made from a 79 nm-thick AlAs and a
67~nm-thick GaAs $\lambda/4$ layer. A single layer of
self-assembled (In,Ga)As/GaAs QDs serves as optically active
medium in the center of the resonator, where the vertical electric
field amplitude has an antinode, to maximize the light-matter
interaction in the planar cavity case. The nominal material
composition of the QDs is InAs, but during growth intermixing with
the GaAs barriers occurs. As the precise, position-dependent
material composition is not available, we use the generic term
(In,Ga)As for the QD material. The QD surface density is $\approx
3 \times 10^{10} $cm$^{-2}$. Single pillar microcavities with
different diameters ranging from about 1 to 6~$\mu$m and spaced
400~$\mu$m apart were fabricated by electron beam lithography and
dry etching \cite{rohner97}. Due to this patterning the field
strength is modulated in the cavity plane.

The pillar microcavities were mounted on the cold finger of a
microscopy flow-cryostat allowing for temperature variations down
to 6~K. In time-integrated photoluminescence spectroscopy, a
frequency doubled Nd:YAG laser was used for continuous wave
optical excitation at $\lambda$ = 532 nm. For time-resolved
photoluminescence, a pulsed Ti:Sapphire laser with pulse durations
of $\approx $100~fs was used. The laser beams were focussed onto
the sample by a microscope objective with a focal length of 1.3
cm, by which a spot diameter of about 10 $\mu$m could be reached.
This diameter is larger than that of the largest studied cavity,
so that one can assume homogeneous excitation conditions.

We note, that the laser excitation does not lead to any sizeable
sample heating effects in the microscopy cryostat. This has been
checked by performing time-resolved experiments with the sample
inserted in superfluid helium ($T$ = 2 K) in an optical bath
cryostat. Under otherwise identical excitation conditions, the
same behavior is observed for the micropillar photoluminescence
kinetics as the one described in Section III.

In particular, the measured time evolution of the PL depends
strongly on the excitation power. In the following we give the
average power density $P_{exc}$, which is connected to the energy
per laser pulse $J_{pulse}$, focussed into a spot with area $A$,
by: $P_{exc} \times A = f \times J_{pulse}$, where $f = $ 75.6 MHz
is the pulse repetition rate. A laser power of 1 mW (corresponding
to a power density of 1.27 kW cm$^{-2}$, which is a typical value
for high excitation in the experiment) is achieved for a pulse
energy of $\sim$ 0.013 nJ. With an excitation energy of 1.4 eV per
electron-hole pair, about 5.8 $\cdot 10^8$ of such pairs could be
created. Further we assume that within the $\lambda$-cavity only
about 0.01\% of the incident laser power is converted into
carriers, which may be trapped in the wetting layer and relax
further towards the QD ground state. This is estimated from
comparing the relative emission intensities of the QDs with that
of bulk GaAs. Distributing 0.01\% of the carriers over the
excitation area leads to an estimate of $n_{eh} \approx 7.5 \times
10^{9}$~cm$^{-2}$ for the carrier density. Carrier densities in
this range have also been used in the numerical analysis, see
Section \ref{sec:theory}.

Some effects have not been considered in this simple estimate,
because they are difficult to quantify. For example, the true
carrier density may be reduced due to above-stopband reflection.
On the other hand, an increase might occur due to reabsorption of
light emitted from GaAs, such as from the substrate. Our cavities
operate in the weak-coupling regime, so that effects known from
strong coupling play no role here. Note that the variations in the
carrier density due to these additional effects are expected to be
small.

Excitation and collection were done through the microscope. After
light collection, the emission was directed into a 0.5~m
monochromator where the signal could be sent either to a charge
coupled devices camera for time-integrated PL studies (used also
for alignment) or to a streak camera for time resolved experiments
with a resolution of $\approx 20$~ps.

\begin{figure}
    \centering
    \centerline{\psfig{figure=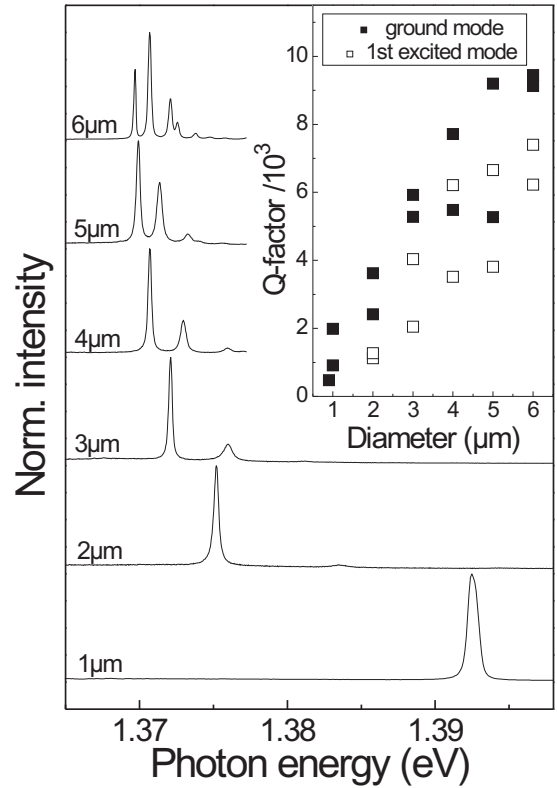,width=\columnwidth}}
\caption{PL emission spectra from single cylindrically shaped
micropillar structures of varying diameters. The inset gives the
quality factors of the two energetically lowest photon modes as
function of the pillar diameter. Solid (open) symbols give data
for the fundamental (first excited) mode. $T$ = 6 K.}
\label{MCfig1}
\end{figure}

In Fig.~\ref{MCfig1} the emission spectra of single pillars with
different diameters are shown, obtained by excitation with the
Nd:YAG laser. For better comparison, the intensity has been
normalized. With decreasing size the energies of the optical modes
shift to higher frequencies. In addition the splitting between the
modes increases strongly. These observations are in accordance
with previous studies of the optical mode spectrum in similar
patterned cavities
\cite{gerard96,reithmaier97,gutbrod98,pelton02,loffler05,daraei06}.
We note that the inhomogeneously broadened QD emission spectrum
has a full width at half maximum of about 30 meV at low
excitation. The emission is centered around 1.38 eV, so that the
QD ensemble represents a light source which provides emission over
the energy range in which the fundamental modes of the studied
optical resonators are located.

We have studied the PL decay of the QDs across their
inhomogeneously broadened emission band (by analyzing the emission
along the plane of an unpatterned resonator and by studying a QDs
reference sample). An initial decay time of 600 $\pm$ 50 ps is
observed at 0.17 kWcm$^{-2}$ with no correlation to the emission
energy. Therefore any cavity size dependence of the carrier
lifetime cannot be related to systematic variations of the dipole
coupling with emission energy due to changes of the QD
confinement.

The inset of Fig.~\ref{MCfig1} gives the pillar diameter
dependence of the quality factors $Q = E / \Delta E $ of the
optical modes. Data for the two lowest confined modes are shown:
in both cases we find a considerable decrease in $Q$ with
decreasing pillar size. While for the 6 $\mu$m diameter pillars
the quality factors are almost 10000, the cavity quality for the
smallest cavities with 1 $\mu$m diameter varies depending on the
resonator from 2000 to below 1000. Also for larger cavity
diameters, sample-dependent variations of the cavity quality are
observed.  Furthermore, the first excited mode has always a
smaller $Q$ than the fundamental mode.  However, one has to be
careful in such a comparison, as an increased line width of the
corresponding emission might arise from a slight mode splitting,
which has been predicted in Ref.~\cite{benyoucef04} for the first
excited mode. Within our experimental accuracy (0.2~meV resolution
of the setup), this mode splitting lies below the mode line width
due to the finite photon lifetime. Independent of the involved
photon mode, the decrease of $Q$ with decreasing pillar size
arises from reduced confinement of the mode for small cavities and
also from an increased importance of surface roughness scattering
at the sidewalls.

\section{\label{Results} Experimental Results}

\begin{figure}
  \centering
  \centerline{\psfig{figure=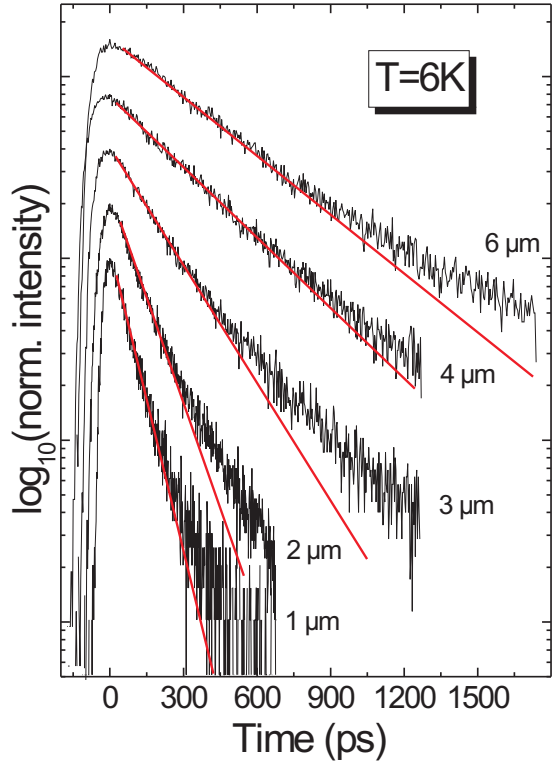,width=\columnwidth}}
\caption{(color online) Low excitation time-resolved PL emission
for micropillars with different diameters. The excitation power
density was 1.3 kWcm$^{-2}$. The decay times corresponding to the
single exponential fits shown by the solid lines are: 400~ps
($6\mu$m), 315~ps ($4\mu$m), 200~ps ($3\mu$m), 110~ps ($2\mu$m),
and 80~ps ($1\mu$m). For clarity, the traces have been shifted
vertically.} \label{MCfig2}
\end{figure}

Figure~\ref{MCfig2} shows the decay of the time-resolved PL for
detection at the energy of the respective fundamental optical mode
of micropillars with different diameters.  The intensity is
plotted on a logarithmic scale. Excitation was done with the
pulsed Tisapphire laser, with the wavelength set to 800 nm,
corresponding to creation of carriers in the GaAs barriers, to
allow for a variation of excitation power and therefore carrier
density over wide ranges. This wavelength is also above the
stop-band of the planar resonator. The used low excitation power
of 1.3 kWcm$^{-2}$ guarantees that the observed PL occurs in the
spontaneous emission regime.

The faster decay for decreasing pillar diameter is mainly a
consequence of the confinement induced enhancement of the vacuum
field amplitude which results in the Purcell effect. The reduction
of mode volume leads to an increase of the Purcell factor, as for
the discussed range of diameters the mode volume decreases faster
than the Q-factors (see inset of Fig.~\ref{MCfig1}). For the decay
of the signal over the first order of magnitude, the deviation
from an exponential decay is rather weak. Straight lines have been
added to fit a T$_1$ time to the initial decay. However, on a
larger scale the decay data clearly reveal a non-exponential
character.

Here we note explicitly that the data cannot be described by
biexponential or stretched exponential decay forms. While such
forms naturally can match the data better than monoexponential
decays, as they involve more fit parameters, they still result in
considerable deviations from the data. Good agreement can
generally only be reached by multiexponential decays involving a
large number of parameters without physical meaning. Biexponential
forms would be appropriate if there were two independent decay
channels, each with a considerable contribution to the emission.
Potential candidates for additional decay channels besides the
excitonic one such as spin-dark excitons, charged excitons, and so
on will be explicitly ruled out by the arguments given below.

\begin{figure}
  \centering
  \centerline{\psfig{figure=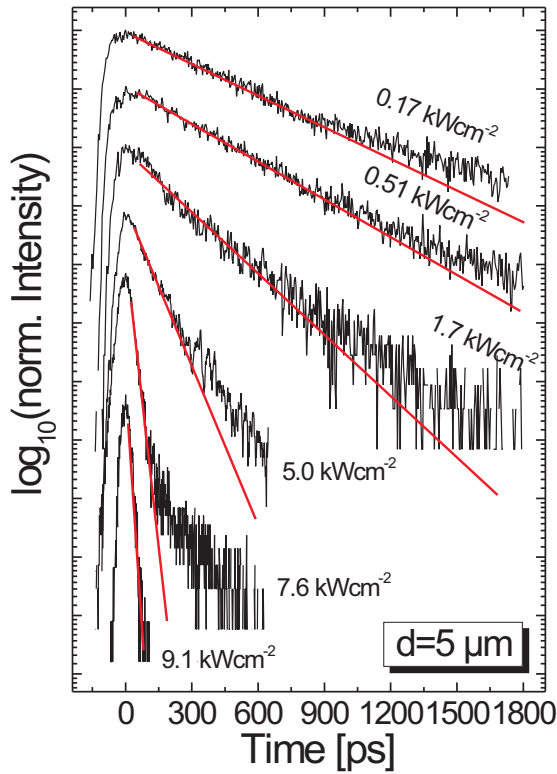,width=\columnwidth}}
\caption{(color online) PL decay curves of a 5 $\mu$m cavity at
different excitation powers. The decay times corresponding to the
single exponential fits shown by the solid lines are: 550~ps
(0.17~kWcm$^{-2}$), 475~ps (0.51~kWcm$^{-2}$), 265~ps
(1.7~kWcm$^{-2}$), 120~ps (5.0~kWcm$^{-2}$), 30~ps
(7.6~kWcm$^{-2}$), and 20~ps (9.1~kWcm$^{-2}$). $T$ = 6 K. For
clarity, the traces have been shifted vertically.} \label{MCfig3}
\end{figure}

To obtain more insight into the emission dynamics, we have varied
the excitation power density $P_{exc}$. Fig.~\ref{MCfig3} shows
the time-resolved emission of a 5 $\mu$m pillar for different
$P_{exc}$ at an excitation wavelength of 800 nm. With increasing
power, the decay becomes generally faster. Even for the lowest
$P_{exc}$, for which we could record a time-resolved signal, no
saturated, power-independent decay is found. For the highest
excitation powers, rise and initial decay become more and more
symmetric with respect to the signal maximum, indicating that the
resonator has been pushed into the stimulated emission regime.
Similar behaviors have been observed for pillars with other
diameters.

First, we need to address the question, to what extent the
observed dependencies regarding cavity size and excitation power
are influenced by non-radiative decay channels, such as traps at
the etched cavity sidewalls. Increasing importance of such traps
with decreasing cavity diameter might also lead to the lifetime
shortening observed in Fig. 2. To analyze this effect we have
performed studies at varying temperatures. For $T > 50$~K thermal
emission out of the QD confinement becomes important, leading to a
drop of PL intensity. At lower $T$, however, the integrated
intensity is constant and also the decay times do not vary with
temperature, which indicates a negligible influence of
non-radiative decay. In this regime, the emitter linewidth $\Delta
\lambda_e$ clearly falls below the cavity mode linewidth $\Delta
\lambda_c$, which is a prerequirement for the observation of the
Purcell effect.

The minor importance of non-radiative decay is also supported by a
variation of the experiment, where the excitation wavelength is
changed to 860 nm, which is slightly above the resonator stop-band
edge, but below the GaAs barrier and into the wetting layer. This
wavelength shift affects carrier capture and relaxation, but
should not influence the cavity size dependence of the decay rate.
However, it would be of importance if non-radiative decay were
relevant, as for excitation above the barrier the photogenerated
carriers may diffuse to the cavity sidewalls, while for excitation
into the narrow wetting layer (which is subject to carrier
localization effects at low temperatures) diffusion to the
sidewalls is strongly hampered. Our data show no change in the
decay rate shortening with decreasing cavity size for the two
excitation conditions. The entirety of these tests allows us to
relate the PL decay time to the radiative decay.

Different reasons for a deviation of the PL emission dynamics from
an exponential decay need to be considered. For very low
excitation power, contributions from exciton complexes such as
biexcitons can be ruled out. However, charged exciton complexes
may be formed due to unintentional background doping, for example.
We have tested the presence of residual charge carriers on a QD
reference sample by Faraday rotation measurements \cite{Awschalom}
in a magnetic field normal to the heterostructure growth
direction. Presence of free carriers would lead to observation of
pronounced spin quantum beats which last longer than the exciton
lifetime. As such beats could not be observed, we can safely
conclude that the vast majority of QDs is undoped.

Further, in the present case the excitation was non-resonant
allowing for fast spin relaxation in the barriers. Therefore not
only spin bright excitons are formed, but also spin dark excitons,
which can recombine radiatively only after a spin-flip. Previous
investigations have shown that, at cryogenic temperatures,
spin-flips are strongly suppressed for carriers in the QD ground
states. For electrons the only viable mechanism seems to be the
hyperfine interaction with the lattice nuclei, while spin-orbit
interaction has been shown to give spin-flip rates in the
kHz-range only, corresponding to much longer time scales than the
ones considered here \cite{BrandesPRB02}. Spin-orbit interaction
is also the only viable mechanism for the holes, but at low
temperatures a two-phonon process is required to induce a spin
flip \cite{KhaetskiiPRL02}. In agreement with these
considerations, exciton spin-flip times have been reported which
are much longer than its radiative decay time
\cite{PaillardPRL01}.

If feeding of the spin-bright exciton reservoir from the reservoir
of dark excitons were important, two time-scales would be
relevant: Apart from the bright exciton decay during times $<$ 1
ns the dark exciton background would decay on time scales of
nanoseconds or longer. From the study of a QD reference sample we
find that no such background can be identified at cryogenic
temperatures. For the time scales of interest it would appear as
contribution to the constant background due to dark counts, which
does not affect the decay analysis, as this background is
subtracted. Thus, also the dark exciton states cannot be thought
of as origin of the non-exponential decay at the used low
temperatures. However, a strong background with decay times in the
few ns range could be observed by raising the temperatures to a
few tens of K, where the thermally induced phonon population can
lead to flip processes through spin-orbit coupling. This is
confirmed by the background decay time shortening strongly with
increasing temperature.

\section{Theoretical Model} \label{sec:theory}

The aim of this section is to outline a theoretical model for the
PL dynamics of QDs, which includes population effects in the
carrier system, the many-body interaction between the carriers as
well as their interaction with the quantized light field within a
microcavity.

QDs are often compared to atomic systems due to the appearance of
localized states with discrete energies.  However, QDs usually
contain many electronic states and excitations involve many
electrons and holes, which are influenced by the Coulomb
interaction. Additional carriers in the WL states contribute to
screening and dephasing which -- together with scattering
processes among the localized carriers -- weakens correlations
between electrons and holes.

This situation differs fundamentally from light-matter interaction
of atomic two-level systems, which are often used for a simplified
analysis of QDs. A two-level model is applicable if the optical
field couples resonantly only to two electronic levels {\em and}
if the excitation involves only a single electron. In this case
the appearance of an electron in the upper state is inescapably
linked to the nonexistence of an electron in the lower state. In
the semiconductor language, electron and hole populations are
fully correlated. As a consequence, the time-resolved PL of
two-level systems shows an exponential decay.

To describe the QD PL, we use the SLE for a system consisting of
interacting charge carriers and a quantized light field, which has
previously been applied to quantum wells \cite{Jahnke1998}. The
SLE describe the coupled dynamics of the electron- and
hole-population, $f^{e,h}_\alpha$, the generalized photon
population, $\langle \op{b}^\dagger_q \op{b}_{q'} \rangle$, and
the photon-assisted polarization, $\langle \op{b}^\dagger_q
\op{h}_\alpha \op{e}_\alpha \rangle$, in the incoherent regime:
\begin{align}
\label{population} i\hbar\frac{d}{d t}f_\alpha^{(e,h)}
\Big|_{\text{opt}}  = & \;
2 i {\rm Re}{\sum_q g^{*}_{q\alpha}\langle \op{b}_q^\dag \op{h}_\alpha \op{e}_\alpha\rangle}, \\
i\hbar\frac{d}{d t}\langle \op{b}_q^\dag \op{b}_{q'} \rangle  = &
\;\label{photons}
\hbar(\omega_{q'}-\omega_q^*)\mean{\op{b}^\dag_q\op{b}_{q'}} \\
\nonumber &\;-\sum_\alpha\left(g_{q'\alpha}^{*}\langle
\op{b}_q^\dag \op{h}_\alpha
\op{e}_\alpha\rangle-g_{q\alpha}\langle \op{b}_{q'}^\dag
\op{h}_\alpha \op{e}_\alpha\rangle^*\right), \\\nonumber
\label{polarisation} i\hbar\frac{d}{d t}\langle \op{b}_q^\dag
\op{h}_\alpha \op{e}_\alpha \rangle  = & \;
(\tilde{\epsilon}_\alpha^e+\tilde{\epsilon}_\alpha^h-\hbar\omega_q^*)\langle \op{b}_q^\dag \op{h}_\alpha \op{e}_\alpha \rangle\\
& - (1-f_\alpha^e-f_\alpha^h) \sum_{\beta} V_{\alpha \beta}
\mean{\op{b}_q^\dag \op{h}_\beta \op{e}_\beta} \\\nonumber & -
(1-f_\alpha^e-f_\alpha^h) \sum_{q'} g_{q'\alpha}
\mean{\op{b}_q^\dag \op{b}_{q'}} \\\nonumber & + g_{q\alpha} (
f_\alpha^e f_\alpha^h+ \Omega^{{cor}}_{q,\alpha} ) \; .
\\\nonumber
\end{align}
Here, $\op{e}^\dag_\alpha$, $\op{e}_\alpha$ and
$\epsilon^e_\alpha$  denote the creation- and
annihilation-operators and the single-particle energy of an
electron in state $\phi^e_\alpha(\vec{r})$. The corresponding
quantities for the holes are $\op{h}^\dag_\alpha$,
$\op{h}_\alpha$, $\epsilon^h_\alpha$, and
$\phi^h_\alpha(\vec{r})$. The states for charge carriers are
either delocalized WL or localized QD states. The operator
$\op{b}^\dag_q$ ($\op{b}_q$) creates (destroys) a photon in the
optical mode $q$, which is characterized by the complex resonance
frequency $\omega_q$ and the transversal mode-pattern
$\vec{u}_q(\vec{r})$. The light-matter coupling is determined by
$g_{q\alpha}\propto \int d^3 r \phi^{e*}_\alpha(\vec{r})
e\vec{r}\vec{u}_q(\vec{r}) \phi^{h}_\alpha(\vec{r})$. The
exchange-Coulomb matrix elements are denoted by $V_{\alpha \beta}$
and the single-particle energies including Hartree-Fock
renormalizations are given by $\tilde{\epsilon}_\alpha^{(e,h)} $.
The population changes due to scattering are treated in a
relaxation time approximation.

Eqs.~(\ref{population}) and (\ref{photons})  show that the
dynamics of the carriers and photons are driven by the photon
assisted polarization, which is governed by
Eq.~(\ref{polarisation}). The first line of
Eq.~(\ref{polarisation}) describes the free evolution, the second
line is responsible for excitonic resonances in the spectrum and
the third line describes stimulated emission or absorption. The
last line contains the source term due to spontaneous emission,
which dominates over the stimulated term for weak excitations.
Nevertheless, in high-Q microcavities, the reabsorption of photons
can modify the results for the time-resolved emission even for
weak excitation. On the Hartree-Fock level (corresponding to
uncorrelated carriers) the source term is given by $g_{q\alpha}
f_\alpha^e f_\alpha^h $. Correlations due to Coulomb interaction
of carriers are included in $\Omega^{{cor}}_{q,\alpha}$ and are
evaluated on singlet level \cite{JahnkePhysStatSol}. We
investigated the influence of higher-order  contributions (doublet
level), as they are discussed in \cite{hoyer2003} for a quantum
well, and find that they play no important role in the presence of
the feedback provided by the cavity.

To gain additional insight into the physics described by the
SLE, we use -- only for the following discussion in this paragraph -- some
simplifications:  We disregard stimulated emission/absorption and neglect the
Coulomb interaction. The adiabatic solution of
Eq.~(\ref{polarisation}) then yields for the population dynamics
\begin{equation}
  \frac{d}{d t}f_\alpha^{(e,h)} \Big|_{\text{opt}}
  = - \frac{f^e_\alpha f^h_\alpha}{\tau_{\text{sp}}}~,
\end{equation}
where $\tau_{\text{sp}}$ is the time-constant for spontaneous
emission. The HF contribution to the source term $f^e_\alpha
f^h_\alpha$ clearly leads to a non-exponential decay. Furthermore,
the rate of decay depends on the carrier density and is higher for
larger population. For the calculations presented below, none of
the mentioned simplifications have been made: stimulated
emission/absorption are included and Coulomb effects such as
corrections to the HF factorization of the source term of
spontaneous emission are considered.

The discussed experiments can be used to distinguish between the
two regimes of fully correlated electron-hole pairs, leading to an
exponential decay of the PL, and partially correlated carriers
with a non-exponential PL decay. As one is typically interested in
the luminescence decay  over a ns time-scale rather than the
initial excitation and relaxation of the system (which takes place
on a ps time-scale), we focus here on the dynamics that occurs
after the system has been excited and the coherent polarization
has decayed due to dephasing. In this incoherent regime we can use
a Fermi-Dirac distribution of carriers in the WL and QD states and
zero photons in the cavity as our initial conditions and evolve
the system according to
Eqs.~(\ref{population})--(\ref{polarisation}).

In order to correctly account for the effects of the size
distribution of  the QDs (different transition energies), their
spatial distribution inside the cavity, and the different dipole
orientations  (different coupling matrix elements), it is not
sufficient to analyze the equations for a single dot with averaged
properties. Instead it is necessary to solve the SLE for an entire
ensemble of different QDs. For our calculation we take only a
fraction of the total number of QDs with a transition frequency
close to that of the relevant cavity modes. Therefore, the
effective density of QDs resonantly interacting with the cavity
mode is assumed to be $3\times 10^9\,\mathrm{cm}^{-2}$,
distributed in an interval of approximately 1.5$\,$meV.

The individual QDs are modeled  with a harmonic confinement
potential in the WL plane and a step-like confinement in growth
direction. The strength of the harmonic confinement is varied for
different QDs to account for the inhomogeneous broadening
typically observed in this material system. We restrict our
analysis to QDs with $s$- and $p$-shells for electrons and holes.

The transverse mode-pattern and resonance  frequencies
$\omega_q^{\text{res}}$ of the optical modes are calculated using
a three-dimensional transfer-matrix approach  (for details see
\cite{benyoucef04}). The corresponding quality-factors Q are
obtained from the experiment. The complex resonance frequency is
then given by $\omega_q = \omega_q^{\text{res}} ( 1 - i/Q) $.
While it is sufficient to include only one resonant mode for the
smaller pillar, for the larger pillars several modes have to be
taken into account. The coupling between different modes is
neglected, $\langle \op{b}_q^\dag \op{b}_{q'} \rangle \approx
\delta_{qq'}\langle \op{b}_q^\dag \op{b}_{q} \rangle$. Besides the
resonant modes, which are characterized by their large $Q$-values
and pronounced peak structure in a transmission spectrum, there
exists a background contribution from the continuum of leaky
modes. In order to include their influence, we assume that the
background contribution  consists of a fraction of the continuum
of modes of the homogenous space. The size of this fraction can be
estimated by counting the plane waves that i) either reach the
sidewalls of the micropillar in an angle smaller than the critical
angle  of total internal reflection, or ii) have a momentum
component $k_{||}$ along the pillar axis that lies outside the
stopband of the DBR and can therefore immediately escape from the
cavity.

\begin{figure}
  \centering
  \centerline{\includegraphics[width=\columnwidth]{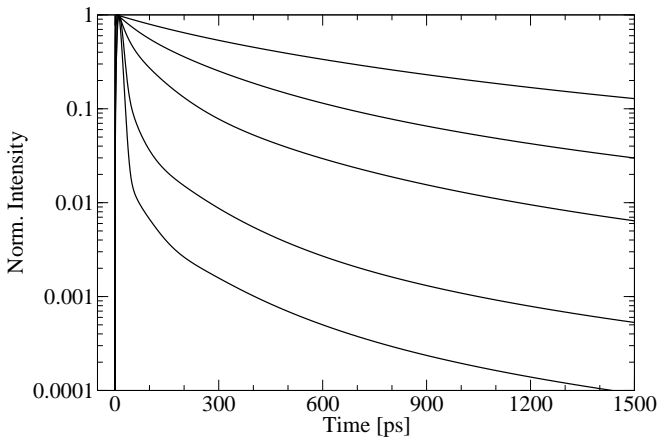}}
  \caption{Calculated PL for an ensemble of
QDs in a 6~$\mu$m diameter pillar microcavity with initial
carrier densities from $1\times 10^{9}$\,cm$^{-2}$ to $5\times 10^{9}$\,cm$^{-2}$ in equidistant steps from top to bottom. For better comparison the results are
normalized.} \label{theory1}
\end{figure}

In Fig.~\ref{theory1} we show the number of photons in the
fundamental mode leaving the cavity per unit time. Different
initial carrier densities are used to model the variation of the excitation power in the
experiment. The non-exponential decay of the
signal is clearly evident. Furthermore, the rapidness of the decay
strongly depends on the initial carrier density, which corresponds to different pump intensities in the experiment. This shows that
it is not meaningful to introduce a decay time that depends only
on the photonic density of states without including the influence of the carrier
system. Instead, a thorough analysis of time-resolved PL signals
has to take  both the carrier system and the photonic system into
account.

It should be noted that for strong optical fields the generated
carrier density depends in a nonlinear manner on the pulse
intensity due to saturation effects. It is not the purpose of the
paper to quantify these optical nonlinearities together with the
subsequent carrier relaxation and to directly connect experimental
pump intensities and the resulting carrier densities. Instead we
focus on the physics of the recombination dynamics and emphasize
the strong dependence of the time-resolved PL decay on the carrier
density in the system.

\begin{figure}
  \centering
  \centerline{\includegraphics[width=\columnwidth]{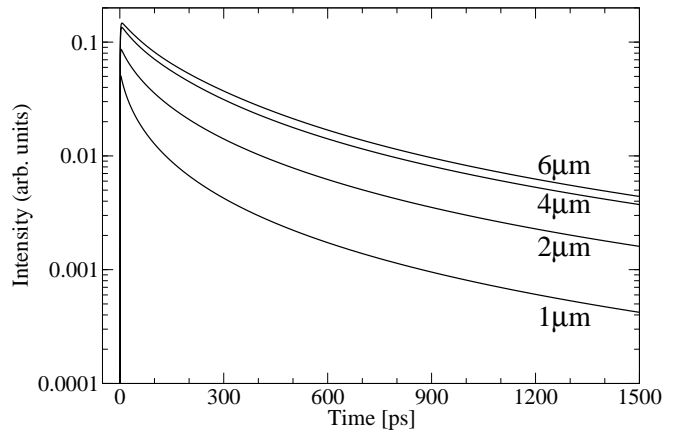}}
  \caption{Calculated PL of QDs in a pillar
microcavity with various diameters for an initial carrier density
of $2 \times 10^{9}\,$cm$^{-2}$.} \label{theory2}
\end{figure}

The calculated PL for fixed initial carrier density but different
diameters of the micropillar cavity is displayed in
Fig.~\ref{theory2}. The smaller pillars show a faster decay in
connection with a larger Purcell factor, as has been discussed in
Sec.~\ref{Results}.  The different heights of the curves can
mainly be attributed to the fact that in larger pillars more
carriers distributed over more QDs take part in the recombination
dynamics.

\begin{figure}
  \centering
  \centerline{\includegraphics[width=\columnwidth]{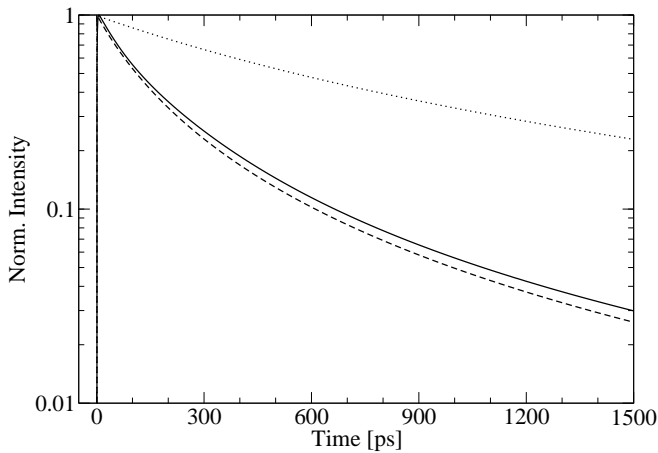}}
  \caption{Calculated PL of an ensemble of QDs with inhomogeneous
broadening (solid line), for identical QDs with maximum coupling
strength and on-resonance transitions (dashed line), and for
identical QDs with averaged coupling strength (dotted). The pillar
diameter is $6\,\mu$m and the initial carrier density is $2\times
10^9\,\mbox{cm}^{-2}$.} \label{theory3}
\end{figure}

Frequently it is argued, that the non-exponential decay observed
in PL measurements stems from a superposition of many exponential
PL signals of various emitters with different cavity positions.
The role of an inhomogeneous distribution of QDs is analyzed in
Fig.~\ref{theory3}. The solid line represents the calculated decay
of the time-resolved PL from an ensemble of QDs with various
cavity positions and fluctuations of the transition energies and
dipole moments (same as in Figs.~4 and 5). For comparison, the
dotted line shows the result for identical QDs with averaged
values for mode-strength, cavity field, transition energy, and
dipole coupling. While the decay remains non-exponential, the
decay rate is strongly underestimated. If identical QDs with
maximum values for mode-strength, cavity field, and dipole
coupling as well as resonant transition energies are assumed
(dashed line), the decay rate is slightly overestimated in the
example with practically the same shape as for the inhomogeneous
QD distribution. This result shows that the QDs with efficient
coupling to the cavity field dominate the emission properties. The
non-exponential character of the decay is only weakly determined
by inhomogeneous distribution effects.

The deviations of the measured from the calculated results close
to time $t=0$ and  in particular the somewhat slower raise of the
PL signal observed in the experiment can be attributed partially
to the fact that the optical carrier generation was not modelled
and that the experimental setup has only a finite time resolution.
Note that we did not adjust the calculations to have a
quantitative agreement between experiment and theory. Too many
parameters, such as size and composition of the QDs, are unknown
in detail.

\section{\label{Summary}Summary}

A microscopic description of the QD emission, based on the
``semiconductor luminescence equations'' including many-body
Coulomb effects, shows the appearance of a non-exponential decay
that is intimately connected with the intensity dependence of the
decay from weak to strong excitation conditions.  The results
explain the time-resolved PL of QDs in pillar microcavities. Other
origins such as contributions from spin-dark excitons, charged
excitons etc. have been ruled out as dominant contribution to the
observations for the presented experiments.

\section{Acknowledgements} This work was support by the research
group 'Quantum Optics in Semiconductor Nanostructures' funded by
the Deutsche Forschungsgemeinschaft. The Bremen group acknowledges
a grant for CPU time at the Forschungszentrum J\"ulich, Germany.

\end{document}